\documentclass[11pt,titlepage]{article}
\usepackage[margin=1in]{geometry}
\usepackage{setspace}
\usepackage[round,numbers,sort&compress]{natbib}
\usepackage{graphicx}
\usepackage{graphics}
\usepackage{dcolumn}
\usepackage{bm}
\usepackage[usenames]{color}

\usepackage{amsmath}
\usepackage{amssymb}

\input epsf

\newcommand{\refeq}[1]{Eq. \ref{#1}}
\newcommand{\sigm}{{\lambda}}
\newcommand{\visrat}{{\chi}}
\newcommand{\out}{{\mathrm{ex}}}
\newcommand{\ins}{{\mathrm{in}}}
\newcommand{\el}{{\mathrm{el}}}
\newcommand{\hd}{{\mathrm{hd}}}
\newcommand{\mem}{{\mathrm{mm}}}
\newcommand{\bE}{{\bf E}}
\newcommand{\bn}{{\bf n}}
\newcommand{\bT}{{\bf T}}
\newcommand{\half} {{\frac{1}{2}}}
\newcommand{\bI}{{\bf I}}
\newcommand{\br}{{\bf r}}
\newcommand{\im}{{\mathrm i}}
\newcommand{\bv}{{\bf v}}
\newcommand{\bnabla}{\nabla}
\newcommand{\Ca}{\mbox{\it Ca}}
\newcommand{\rhat}{{\bf \hat r}}
\newcommand{\brac}[1]{\left(#1\right)}

\title{Electrohydrodynamic model of vesicle deformation in alternating electric fields}

\author{Petia M. Vlahovska\thanks{
            Corresponding author.  Address:
Thayer School of Engineering,
Dartmouth College, 8000 Cummings Hall
Hanover, NH 03755, USA
Tel.: (603) 646-9922
email: petia.vlahovska@dartmouth.edu}\\
Thayer School of Engineering\\
Dartmouth College\\
Hanover, NH 03755, USA
\and   Rub\`en Serral Graci\`a\thanks{
            Current address: Culgi B.V.
P.O. Box 252
2300 AG Leiden, The Netherlands}, Said Aranda, Rumiana Dimova\\
Max Planck Institute of Colloids and Interfaces\\
Science Park Golm\\
14424 Potsdam, Germany
}

\date{\today}
\begin{document}

\maketitle

\abstract{We develop an analytical theory to explain the experimentally-observed morphological transitions of giant vesicles induced by AC electric fields \cite{Aranda:2007}.
The model treats the inner and suspending media as lossy dielectrics, and the membrane as an ion-impermeable flexible incompressible-fluid sheet. The vesicle shape is obtained by balancing electric, hydrodynamic, and bending stresses exerted on the membrane.  Considering a nearly spherical vesicle, the solution to the electrohydrodynamic problem is obtained as a regular perturbation expansion in the excess area.

The theory predicts that stationary vesicle deformation depends on field frequency and conductivity conditions. If the inner fluid is more conducting than the suspending medium, the vesicle always adopts a prolate shape. In the opposite case, the vesicle undergoes a transition from a prolate to oblate ellipsoid at a critical frequency, which the theory identifies with the inverse membrane charging time. At frequencies higher than the inverse Maxwell-Wagner polarization time, the electrohydrodynamic stresses become too small to alter the vesicle's quasi-spherical rest shape. The analysis shows that the evolution towards the stationary vesicle shape strongly depends on membrane properties such as viscosity.
The model can be used to rationalize the transient and steady deformation of biological cells in electric fields.\\
\\
\vspace{1cm}
}

\clearpage

\section{Introduction}

Electric fields are  widely used for cell manipulation.
Weak fields influence cell signaling, wound healing, and cell growth \citep{Zhaoetal,Funk,Voldman:2006, Zimmermann-Neil:1996}. Strong pulsed fields can induce transient perforation of the cell membrane, which enables the delivery of exogenous molecules (drugs, proteins, and plasmids) into living cells \cite{Zimmermann-Neil:1996, Neumann-Sowers-Jordan:1989}.

Biological cells exhibit various frequency-dependent behaviors in AC electric fields: orientation, translation (dielectrophoresis), and rotation.
These phenomena have stimulated considerable modeling effort aimed at
understanding of the physical mechanisms of the interaction of electric fields with cells and tissues.
A common theme among different theoretical
models is the assumption that the cell  is a lossy dielectric particle of {\it{fixed}} shape (a sphere\cite{Chizmadzhev-Kuzmin:1985, Pastushenko:1988, Turcu, JonesTB} or an ellipsoid\cite{Gimsa-Wachner:1999, Dolinsky-Elperin:2006}).
For example, the orientation of cells  can be predicted
by considering the torque on an ellipsoid due to the effective dipole moment induced by the electric field \cite{Schwan:1992, Jones:2003}; the dipole based theory has been successfully applied to  interpret electro-orientation of erythrocytes \cite{Miller-Jones:1993}.

Cells, however,  are {\it soft} objects, which deform when subjected to electric fields.
The cell membrane  plays a critical role in this process. A number of studies have focused on the  membrane shear   elasticity  because of
the interest in the mechanics of the red blood cell \cite{Engelhardt-Sackmann:1988, Bryant-Wolfe:1987, Pawlowski:1992}.  The  lipid bilayer is  the main structural component of the cell membrane, yet,
the electrodeformation of closed pure lipid bilayer membranes (vesicles) has been considered only to a limited extent  \cite{Helfrich:1974, Kummrow-Helfrich:1991, Winterhalter-Helfrich:1988}. There is increasing interest in this problem, particularly in relation to electropermeabilization \cite{Teissie:1981, Sukhorukov:1998, Isambert:1998, Sens-Isambert:2002, Lacoste:2007, Ambjornsson:2007, Krassowska:2007}. Recent experiments have shown  that vesicle behavior in electric fields exhibits peculiar features. Vesicles subjected to a direct-current-electric pulse  can deform into elliptical \cite{Riske-Dimova:2005} or  cylindrical shapes \cite{Riske-Dimova:2006}.
Alternating-current-electric fields deform vesicles into prolate or oblate ellipsoids depending on the frequency and the conductivities of the interior and suspending fluids~\cite{Rumy:2006, Dimova-Aranda:2007, Aranda:2007}.

The physical mechanisms responsible for vesicle electrodeformations are not fully understood at present time.  A prolate shape can be explained by
the  electric pressure pulling the vesicle at the poles.
However, the  oblate shapes remain an open problem, in particular, the fact that they are observed only when the conductivity ratio of inner and outer fluids is less than unity \cite{Dimova-Aranda:2007}.
Attempts to explain the oblate shapes have been made
\cite{Mitov:1993}.  Peterlin et al. \cite{Peterlin:2007} showed that anisotropy in the dielectric constant of the membrane could lead to oblate shapes, but this model does not account for
the observed dependence on the conductivity ratio.
Hyuga et al. \cite{Hyuga:1991a} realized that  the  fluid environment is not just  a passive milieu and that the dynamic coupling between  changes in membrane shape and motion in the surrounding fluids is important. However, the fluid flow in their model was described  by an ad hoc equation that includes fluid acceleration. On the contrary,  fluid motion at the micro-scale is in the viscous Stokes regime, where friction effects dampen fluid acceleration \cite{Nelson}. Thus, the validity of their analysis is questionable.

In this paper we  develop
a  model for vesicle dynamics in electric fields.
The transient vesicle deformation is determined by evaluating the  forces
exerted on the membrane  \cite{Sauer}.
Our theory
builds on the large amount of research devoted to the  electrohydrodynamics of drops \cite{Taylor:1966, Melcher-Taylor:1969, Saville:1997}. Drops can adopt oblate  shapes because the electric field causes continuous fluid flow, which pushes fluid toward the equator. The physical mechanism behind the electrohydrodynamic flow is the following: If the fluids are leaky dielectrics, i.e., possess  finite conductivity,
free charges  accumulate at the drop interface. The electric field acting on these changes creates a tangential electric force, which drags the fluids into motion (see Figure 1).

Notwithstanding the qualitative similarity between drops and vesicles, the extension of the ``leaky dielectric'' model  from drops to vesicles is not a straightforward task because the mechanics of lipid membranes is far more complex than the mechanics of fluid-fluid interfaces.  There are two major challenges.
First, the lipid membrane
is essentially an insulating shell impermeable to ions. When an electric field is applied,  charges accumulate on both sides of the bilayer and the  vesicle acts as a charging capacitor.
Second, since the lipid bilayer contains a fixed number of molecules the membrane is incompressible.  Under stress, the membrane develops tension, which
adapts itself to the forces exerted on the membrane in order to keep the local area constant. At steady state the gradients in tension counteract the tangential electric force and the electrohydrodynamic flow stops.
Our approach rigorously accounts for these phenomena.

The paper is organized as follows: Section \ref{sec1} describes the physical model and formulates the governing equations, Section \ref{sec2} outlines the solution and discusses the frequency dependence of the electric stresses, and  Section \ref{sec3}  shows the theory predictions for the vesicle shape as a function of frequency, conductivity ratio and other physical parameters of the system such as membrane viscosity.

\section{The model}
\label{sec1}

\subsection{The physical picture and characteristic time scales}
Let us consider a
 vesicle with no net charge  formed  by a
 membrane with  conductivity $\sigm_{\mem}$, dielectric constant $\epsilon_\mem$, and surface viscosity
$\eta_\mem$.  The bilayer thickness is about $h\sim~5nm$, thus  on the length scale of a cell-size vesicle (radius $a \sim~10\mu m$) the bilayer membrane can be regarded as a two-dimensional surface with capacitance $C_m=\epsilon_\mem/h$ and conductivity $G_m=\sigm_{\mem}/h$. The  vesicle is filled with a fluid of viscosity
$\eta_\ins$, conductivity $\sigm_{\ins}$, and dielectric constant $\epsilon_\ins$, and suspended in a different
fluid characterized by $\eta_\out$,  $\sigm_{\out}$, and  $\epsilon_\out$. The physical properties of the fluids and the membrane are assumed to be frequency-independent.

The vesicle is subjected to a uniform AC electric field with an amplitude $E_0$
\begin{equation}
\label{external field}
{E}^\infty=E_0 \cos(\omega t),
\end{equation}
where $\omega$ is the angular field frequency and $t$ is the time.
Free charges move
under the action of the electric field.
The density of free charges in the bulk fluids
decays on a time scale \cite{Melcher-Taylor:1969, Saville:1997}
\begin{equation}
\label{tc}
t_{c}=\frac{\epsilon_i}{\sigm_i}\,,\qquad i=\ins, \out
\end{equation}
where $\epsilon$ and $\sigm$ denote the absolute permittivity and conductivity of the fluid.   Hence, for frequencies $\omega<t_{c}^{-1}$,  free charges are present only at
 boundaries that separate media with different electric properties.
The rate of accumulation of charges at the  interface of a macroscopic object, e.g., a sphere, is given  by the Maxwell-Wagner polarization time \cite{JonesTB}
\begin{equation}
\label{tMW}
t_{MW}=\frac{\epsilon_\ins+2\epsilon_\out}{\sigm_\ins+2\sigm_\out}\,.
\end{equation}

The electric field acts on the free charges at the interface  and  gives rise to a force, which is tangential to the interface. In the case of a simple fluid-fluid interface, e.g., a  drop,
only a hydrodynamic force can balance the shearing electric force. As a result, the fluids are set in continuous motion, the so called electrohydrodynamic (EHD) flow \cite{Taylor:1966}. The EHD flow is characterized by a time scale, which  corresponds to the inverse of the shear rate imposed by the tangential electric stress
\begin{equation}
t_{\el}=\frac{ \eta_\out}{\epsilon_\out E_0^2}\,.
\end{equation}
In the case of drops, the flow inside is toroidal with a direction either from or towards the poles depending on the surface charge distribution as illustrated in Figure \ref{fig: toroid}.

The membrane represents a more complex boundary compared to fluid-fluid interfaces. First, it  is an insulating shell and charges accumulate on both the inner and outer surfaces, as illustrated in Figure~\ref{sketch}.
Hence, a vesicle of radius $a$ acts as spherical capacitor that charges on a time scale given by \cite{Schwan, Kinosita:1988}
\begin{equation}
\label{tcap}
t_{cap}=a C_m \left(\frac{1}{\sigm_\ins}+\frac{1}{2\sigm_\out}\right)\,.
\end{equation}
Second, the membrane mechanics is governed by resistance to bending.
A distortion of the membrane shape
relaxes on a  time scale
\begin{equation}
t_{\kappa}=\frac{\eta_\out a^3}{\kappa}\,,
\end{equation}
where $\kappa$ is the bending modulus.
The curvature relaxation depends on the viscosity of the surrounding fluids, because the  membrane
has to move fluid in order to return to its preferred configuration.

It is instructive to estimate the magnitude of the  characteristic time scales involved in vesicle electrodeformation. Typical experimental conditions
involve fluids with conductivities in the range  $\sigm\sim10^{-3}-10^{-4} S/m$ and electric fields of the order of  $E\sim 1kV/cm$ \cite{Kummrow-Helfrich:1991, Mitov:1993, Riske-Dimova:2006, Dimova-Aranda:2007, Aranda:2007}.  In physiological conditions, e.g., blood,  the internal conductivity of an erythrocyte is much higher  $\sim 0.5 S/m$ \cite{Miller-Jones:1993}. The typical size of a vesicle or cell is $a\sim 10\mu m$. The inner and outer fluids are essentially
water: viscosity $\eta\sim10^{-3} Pa.s$,and density  $\rho\sim1000 kg/m^3$.
The membrane capacitance is $C_m \sim 10^{-2} F/m^2$ \cite{Needham-Hochmuth:1989} and bending rigidity $\kappa\sim10^{-19} J$. Therefore, for vesicles, we estimate the basic charging time and the Maxwell-Wagner polarization time  $t_c\sim t_{MW} \sim 10^{-7}s$, the membrane charging time  $t_{cap} \sim 10^{-3} s$,  the electrohydrodynamic time  $t_\el \sim 10^{-3} s$, and the bending relaxation time  $t_\kappa \sim 10 s$.

We see that vesicle dynamics in electric fields involves processes that occur on very different times scales. Vesicle deformation takes place concurrently with fluid motion. The electric field adjusts to a new boundary configuration much faster than the fluid moves, because  conduction (and hence charge redistribution) is fast, $t_{MW}\ll t_{\el}$. Hence, the electric field depends only on the instantaneous vesicle shape; it is quasi-static. The flow time scale is comparable to the capacitor charging time, $ t_{\el}\sim t_{cap}$. The interplay between these two time scales is responsible for the observed dynamics of vesicles in electric fields \cite{Aranda:2007}.

\subsection{Governing equations}

In essence, the leaky dielectric model consists of conservation of current, which obeys Ohm's law, and the Stokes equations to describe fluid motion \cite{Saville:1997}.
Charges carried by conduction  accumulate at interfaces, and bulk phases become charge-free on a very fast time scale given by $t_c$, \refeq{tc}.
Accordingly, electromechanical coupling occurs only at boundaries.

\subsubsection{Electrohydrodynamic problem}
\paragraph{Electric field:}
In the absence of bulk charges, the electric potential, $\Phi$, for a quasi-static electromagnetic field is the solution of the Laplace equation
\begin{equation}
\label{Maxwell eqs}
\nabla^2 \Phi=0\,,\qquad \bE=-\nabla \Phi  \,.
\end{equation}
The membrane acts as a capacitor.
 Accordingly, the potential  undergoes a jump across the interface
\begin{equation}
\label{potential}
\Phi^\ins-\Phi^\out= \Delta\Phi(\omega,t)\, \quad \mbox{at}\quad r=r_s ,
\end{equation}
where $r=r_s$ denotes the position of the interface in a  coordinate system
centered in the vesicle (see Fig.~\ref{sketch}).
The relation between the transmembrane potential and the membrane capacitance depends on geometry. The spherical shell  is a widely used model for cells and vesicles \cite{Schwan, Turcu}, although a spheroidal geometry has also been  considered \cite{Gimsa-Wachner:1999}.
Free charges at the interface cause discontinuity in the normal component of the displacement vector
\begin{equation}
\label{normal balance}
\bn\cdot \left[\epsilon_\out\bE^\out-\epsilon_\ins\bE^\ins\right]=Q(\omega,t)\,\quad \mbox{at}\quad r=r_s,
\end{equation}
where $\bn$ is the outward unit normal vector and $Q$ is the free charge density.
Neglecting effects of charge convection  along the surface by fluid motion, the conservation of electric currents
at the interface requires that
\begin{equation}
\label{charge conservation:AC}
\bn\cdot \left[\sigm_\out\bE^\out-\sigm_\ins\bE^\ins\right]=-\frac{\partial Q}{\partial t} \,\quad
\mbox{at}\quad r=r_s .
\end{equation}
The forces  due to an electric field $\bE$
are  calculated from the Maxwell stress tensor
\begin{equation}
\label{Maxwell stress}
\bT^{\el}=\epsilon\left(\bE\bE-\half\bE^2 \bI\right)\,,
\end{equation}
where $\bI$ denotes the unit tensor.
A harmonic electric field can be written as
\begin{equation}
E \cos(\omega t)
=\half \left(\bE+\bE^*\right)\,.
\end{equation}
It gives rise to a non-zero time-averaged component of the Maxwell stress tensor
\begin{equation}
\label{Max stress}
\bT^\el_s(\omega)=\frac{1}{4}\left[\bE \bE^*+\bE^* \bE-|E|^2 \bI\right]\,,
\end{equation}
which is responsible for the steady deformation of the vesicle.

All electric variables (electric field, potential, charge density) vary harmonically with time
$u(\br, t, \omega)=\bar u(\br,\omega)\exp(\im \omega t)$. Hence, hereafter unless specifically
stated, we will always refer to the amplitude of an electric variable, $\bar u(\br,\omega)$, and we will
omit the bar for convenience.

\paragraph{Hydrodynamic field:}

Vesicle deformation is accompanied by motion in the surrounding fluids.
The  fluid velocity, $\bv$, and pressure, $p$, inside
  and outside the vesicle are described by the Stokes equations \cite{Nelson, Leal:1980}
\begin{equation}
\label{Stokes equations}
\frac{\partial \bv}{\partial t}=\bnabla \cdot\bT^\hd \, , \quad \bnabla\cdot\bv=0\,,
\end{equation}
where the  bulk hydrodynamic stress is
\begin{equation}
\label{stress definition}
\bT^\hd=\textstyle -p\bI+\eta[\bnabla\bv+(\bnabla\bv)^\dagger]\,,
\end{equation}
where  the superscript ${\dagger}$ denotes transpose.

\refeq{Stokes equations} is a simplified version of the more general Navier-Stokes equation. First,  inertial effects are neglected because at the length-scale of the cell water is effectively very viscous. Second, the bulk stress has no contribution from the electric field because there are no  free bulk charges.
Moreover, the unsteady term $\partial \bv/\partial t$ can be neglected if the diffusion of momentum, $t_v=a^2 \rho/\eta$,
is faster that the changes in the electric field, i.e., $\omega<t^{-1}_v$ \cite{Sozou:1972}.  The linearity  and quasi-steadiness of the Stokes equations, and the decoupling of the electric and hydrodynamic equations in the bulk greatly simplify the solution of the problem.

Far away from the vesicle, the fluid is at rest and the flow field vanishes, $\bv^{\out}\rightarrow 0$. Velocity is
continuous across the interface
\begin{equation}
 \bv^{\ins}=\bv^{\out}\equiv\bv_s  \quad \mbox{at}\,\quad r=r_s\,.
\end{equation}
The interface moves with the fluid \cite{Leal:1992}
\begin{equation}
\label{interface evolution}
\frac{\partial r_s}{\partial t}=\bv_{\mathrm{s}}\cdot\bn\,.
\end{equation}


\paragraph{Electromechanical coupling:}
The vesicle shape  is determined by the balance between electric, hydrodynamic, and membrane tractions (surface force densities)
 at the interface $r=r_s$
\begin{equation}
\label{stress balance}
\bn\cdot[(\eta_\out\bT^{\hd,\out}-\eta_\ins\bT^{\hd,\ins})+(\epsilon_\out \bT^{\el,\out}-\epsilon_\ins\bT^{\el,\ins})]
={\bm{\tau}}^{mem} \,,
\end{equation}
where flexoelectric bending of the lipid bilayer is neglected \cite{Petrov:review, Raphael:2000}.
For example, at rest, when the electric field is off,  \refeq{stress balance} reduces to the familiar Euler--Lagrange equation \cite{Seifert:1999}, which states that there can be a  jump in the hydrostatic pressure across a membrane due to membrane tractions
\begin{equation}
\label{Euler-Lagrange}
p^\ins-p^\out=2\sigma H -\kappa \left[4H^3-4KH+2\nabla_s^2 H\right]\,.
\end{equation}
where $\kappa$ is the  bending rigidity, $H$ and $K$ are the mean and Gaussian curvatures.
In the next section we discuss the membrane stresses in more detail.

\subsubsection{Membrane mechanics}

The pure lipid membrane
consists of two sheets of lipid molecules. The molecular thickness
imparts resistance to bending.
Within the framework of the minimal model \cite{Seifert:1997},
the bending resistance gives rise to a surface force density
\begin{equation}
\label{interfacial stress}
{\bm{\tau}}^\kappa=-\kappa \left(4H^3-4KH+2\nabla_s^2 H\right)\bn\,.
\end{equation}
The surface gradient operator
is defined as $\bnabla_{\mathrm{s}}=\bI_\mathrm{s}\cdot\bnabla$,
where the matrix $\bI_\mathrm{s}=\bI-\bn\bn$ represents a surface projection.

The membrane leaflets consist of fixed number of lipids, which are optimally packed with fixed area per lipid (under moderate stresses).
As a result, a membrane element
only deforms but can not change its area. Under stress, the membrane develops tension, which
adapts itself to the forces exerted on the membrane in order to keep the local and total area constant. Hence, the tension is non-uniform along the interface and varies with forcing.
The membrane tension gives rise to  surface force density
\begin{equation}
\label{tension stress}
{\bm{\tau}}^\sigma=2\sigma H\bn-\bnabla_s \sigma \,.
\end{equation}
where $\sigma$ denotes the local membrane tension

Lipid molecules are free to move within
the monolayer, and therefore, in contrast to solid--like polymerized
membranes, the lipid bilayer membrane is fluid with a zero shear-elastic modulus \cite{Rumy:2006}.
How fast the membrane flows, however,  depends on the rate-of-deformation. The viscous stresses developing in the flowing membrane are
\begin{equation}
\label{mem visc stress}
{\bm{\tau}}^\eta=\eta_\mem[\nabla_s\bv_s+(\nabla_s\bv_s)^\dagger]\,.
\end{equation}
The membrane viscosity of lipid bilayers is relatively low, $\eta_\mem \sim 10^{-9} N s/m $, and its effects are usually negligible. However,
polymersomes   can  have very viscous membranes characterized by large membrane viscosity $\eta_\mem\sim 10^{-6} N s/m$ \cite{Discher:2006, Rumy:2002}.

\subsection{Dimensionless parameters}
\label{nondimen}

It is more convenient to describe the problem in terms of non-dimensional parameters. Casting equations in dimensionless form helps show the generality of application to a broad class of situations rather than just one set of dimensional parameters.

Henceforth, bending stresses and tension are normalized by $\kappa/a^2$; all other quantities are
rescaled using $\eta_\out$, $\epsilon_\out$, $\sigm_\out$, $a$, and $E_0$.
The fluid velocity scale is $v_0=\epsilon_\out E_0^2a/\eta_\out$. The electric and viscous stresses are rescaled by $\epsilon_\out E_0^2$. Time and frequency are nondimensionaized with the  basic charging time $t_c=\epsilon_{\out}/\sigm_{\out}$.

The electric capillary number compares the shape-preserving bending stresses to the shape-distorting electric stresses,
\begin{equation}
\Ca=\frac{t_{\kappa}}{t_{\el}}=\frac{\epsilon_\out E_0^2 a^3 }{\kappa}\,.
\end{equation}
The other relevant parameters are the ratios of the electric properties of inner and outer fluid
\begin{equation}
\Lambda=\frac{\sigm_\ins}{\sigm_\out}\,,\quad S=\frac{\epsilon_\ins}{\epsilon_\out}
\end{equation}
and the viscosity ratio
\begin{equation}
\label{visrat}
\visrat=\frac{\eta_\ins}{\eta_\out}\,.
\end{equation}
The dimensionless membrane conductivity and capacitance per unit area are
\begin{equation}
\label{GmCm}
G_m=\frac{\sigm_\mem}{x \sigm_\out }\,, \quad C_m=\frac{\epsilon_\mem}{x\epsilon_\out}\,,
\end{equation}
where  the dimensionless membrane thickness is $x=h/a$.
The importance of membrane viscous stresses  is reflected by the magnitude of the surface viscosity parameter
\begin{equation}
\label{visratS}
\visrat_s=\frac{\eta_\mem}{\eta_\out a}\,.
\end{equation}

 The surface viscosity parameter for a lipid vesicle is relatively small, $\visrat_s \sim 1$, but for polymersomes can be quite large, $\visrat_s \sim 10^3$.

 We estimate that $\Ca\sim 10^{3}\gg 1$ from the typical values discussed at the end of Section \ref{sec1}.1. The dielectric constants and viscosity ratios are $S, \chi \sim 1$.  In physiological environments and biological applications, the conductivity ratio can vary between $10^{-3}$ and 100.

\section{Solution for small deformations}
\label{sec2}

In a coordinate system centered at the vesicle,
the radial position  $r_{\mathrm{s}}$ of the vesicle
interface is described by
\begin{equation}
\label{perturbation of shape}
r_{\mathrm{s}}=1+f(\theta\,, \phi)\,,
\end{equation}
where $f$ is the deviation of vesicle shape from a sphere.
For a nearly spherical vesicle, $f \ll 1$.
In this case, the exact position of the interface
is replaced by the surface of a sphere of equivalent volume,
and all quantities that are to be evaluated at the interface of the
deformed vesicle are approximated using a Taylor series expansion. The solution for electric and flow fields is derived as a regular perturbation expansion in some small parameter, e.g., the excess area.

In this study we perform the leading order analysis. Accordingly, the electric and hydrodynamic fields are evaluated about a sphere. First, we determine the electric field and the electric tractions (surface force density) exerted on the membrane. Second, we determine the hydrodynamic tractions needed to satisfy the force balance \refeq{stress balance} and the corresponding velocity field. Finally, we use the kinematic condition \refeq{interface evolution} to find the shape evolution.

In  \refeq{perturbation of shape}, the function $f$ representing the perturbation of the vesicle shape depends only on angular coordinates.  Thus, it is expanded into series of scalar spherical harmonics
$Y_{jm}$ given by \refeq{normalized spherical harmonics} in Appendix \ref{Harmonics}
\begin{equation}
\label{expansion of shape in harmonics}
f=\sum_{j=2}^{\infty}\sum_{m=-j}^j f_{jm} Y_{jm}\,,
\end{equation}
Solutions for the electric field are  growing and decaying harmonics which derive from $\nabla(r^jY_{jm})$ and $\nabla(r^{-j-1}Y_{jm})$.
The uniform applied electric field along the $z$-direction, defined by \refeq{external field},  is described by the  $j=1$ harmonic
\begin{equation}
\label{Einf}
\bE^\infty= d^\infty\nabla{\left(rY_{10}\right)}\,,
\quad d^\infty=\sqrt{\frac{4 \pi}{3}}
\end{equation}
Accordingly, the induced electric field has $j=1, m=0$ symmetry.

\subsection{Electrostatic field and stresses}

The model for the electric field is based on the classic works by Schwan and coworkers \cite{Schwan}. They have shown that
an external AC electric field induces a potential  across the membrane of a spherical shell
\cite{Grosse-Schwan:1992},
\begin{equation}
\label{potential2}
\Delta \Phi=V_m(\omega)\cos\theta\,,
\end{equation}
where
\begin{equation}
\label{potential3}
V_m(\omega)=\frac{3}{2}\frac{1}{1+(G_m+\im \omega C_m)(1/\Lambda+2)}\,.
\end{equation}
The transmembrane potential is very sensitive to the membrane thickness. Figure \ref{fig:potential}.a  illustrates the variation of the transmembrane potential with frequency  for a vesicle with a fixed size and two values of the membrane thickness, corresponding to a giant unilamelar lipid vesicle and a polymersome.   For a simple fluid-fluid  interface (a drop, i.e., $x=0$), the transmembrane potential is zero.

The electric tractions exerted on the membrane have radial and  tangential components
\begin{equation}
\label{el tractions}
{\bm{\tau}}^\el=\tau_r^\el\left[1+3\cos(2\theta)\right]\rhat  +\tau_\theta^\el\sin(2\theta){\bm{\hat \theta}}  \,.
\end{equation}
In terms of the electric field, the electric pressure can be written as
\begin{equation}
\tau_r^\el=\half[(E_r^\out)^2-(E^\out_\theta)^2-S((E_r^\ins)^2-(E^\ins_\theta)^2)]\,,
\end{equation}
and the tangential electric force is
\begin{equation}
\label{tangT}
\tau_\theta^\el=E^\out_\theta Q+S E^\ins_r V_m(\omega)\sin\theta\,,
\end{equation}
where we have used the definition of surface charge $Q$ \refeq{normal balance}.
The amplitudes of the electric tractions, $\tau_r^\el$ and $\tau_\theta^\el$, depend only on the electric properties of the media.
Their expressions  are given by \refeq{tn} and \refeq{tt} in  Appendix \ref{el field shell}.

The electric stresses are  complicated functions of the frequency $\omega$ as illustrated in Figures \ref{fig:tractions}.a and \ref{fig:tractions}.b. We can distinguish three regimes:

{\it{Low frequencies, $\omega<\omega_1$:}} In this case, the membrane shields the vesicle interior and the electric field inside zero,
as seen from Figure \ref{fig:potential}.b.
The electric pressure
is positive at the poles, and negative at the equator, thus pulling the vesicle into a prolate shape.  The tangential electric stress is zero everywhere on the surface, because both induced charge and internal electric field are zero.
In contrast, the tangential electric stress at a simple fluid-fluid interface, i.e.,  zero-membrane-thickness,
is non-zero even at low frequencies, see \refeq{tt0}.
The electric pressure changes sign and the tangential electric traction becomes significant above a frequency $\omega_1$ given by \cite{Turcu, Schwan}
\begin{equation}
\label{omega1}
\omega_1=\frac{G_m}{C_m}+\frac{2\Lambda}{C_m(\Lambda+2)}\,,
\end{equation}
which reduces to $1/t_{cap}$, \refeq{tcap}, if the membrane is non-conducting.

{\it{Intermediate frequencies, $\omega_1<\omega<\omega_2$:}} In this frequency window, the membrane capacitor becomes ``short-circuited'' and the vesicle interior participates in the conduction process. The onset of decrease in the transmembrane potential and increase in the interior electric field coincides with the appearance of tangential electric tractions and negative electric pressure,  as seen in Figures \ref{fig:potential} and \ref{fig:tractions}.
The tangential electric stress is mainly due to the free charges on the membrane \cite{Taylor:1966}, see \refeq{tangT}.
Because of the different conductivities of the inner and outer fluids, charges accumulate at different rates
on the  membrane physical surfaces. Charge densities on the  inner and outer membrane surfaces can become imbalanced,
which gives rise to  a non-zero  effective interfacial charge density   as shown in Figure  \ref{fig:charge}.
The effective charge is zero at low frequencies because the membrane capacitor is fully charged,
 having equal charge densities  on the inner and outer membrane surface,  and at high frequencies
 because of insufficient time for interface charging.

{\it{High frequencies, $\omega>\omega_2$}}. The inverse Maxwell-Wagner polarization time, \refeq{tMW}, defines a critical frequency
\begin{equation}
\label{omega2}
\omega_{2}=\frac{\Lambda+2}{S+2}\,,
\end{equation}
above which tangential stress starts to decrease. It vanishes at very high frequencies,
where  all media behave as perfect dielectrics. In this frequency regime, the electric pressure is small, but positive  with magnitude $\sim (S-1)^2/(S+2)^2$, which leads to small prolate deformation.

\subsection{Hydrodynamic field and vesicle deformation}

The stress balance at the interface \refeq{stress balance} shows that the electric tractions need to be compensated by membrane and hydrodynamic forces.  The latter can be found using the general solution for a nearly spherical vesicle subject to an external field of arbitrary symmetry developed in Ref.
\cite{Vlahovska:2007}. Details of the solution are presented  in the Supplementary material.


The vesicle area, $A$,
exceeds the area needed to enclose the volume of the interior fluid, $4 \pi a^2$. At rest, the excess area is redistributed among all shape modes
\begin{equation}
\label{excess area}
\Delta=A/a^2-4 \pi=\sum_{j=2}^{\infty}\sum_{m=-j}^j\frac{(-1)^m}{2} (j-1)(j+2) f_{jm}f_{j-m}\,.
\end{equation}
Therefore, in order to accurately describe  vesicle deformation, in general, we need the evolution equations for all shape modes.
These are derived in \cite{Vlahovska:2007} (see also the Supplementary material) and have the general form
\begin{equation}
\label{ev equation f}
\frac{\partial f_{jm}}{\partial t}=C_{jm}+\Ca^{-1}(\Gamma_1+\sigma_0\Gamma_2)f_{jm}+O\left(f^2\right)\,.
\end{equation}
The first term describes the distortion of the vesicle shape
by  the electrohydrodynamic flow. The term including
$\Ca$ is associated with shape relaxation driven by the membrane
stresses. The coefficients $C_{jm},\, \Gamma_1$, and $\Gamma_2 $
are
 listed in the Supplementary material. The effective tension $\sigma_0$  depends on the vesicle shape, which in turns depends on the applied electric field.

In general, the apparent area of a vesicle, $\bar A$, is lower than its true area, $A$, because of suboptical fluctuations in the shape modes.
For example, a quasi-spherical vesicle at equilibrium is characterized by a zero apparent area, i.e., $\bar \Delta=0$. However, even though the membrane in inextensible, the vesicle can deform and increase its apparent  area due to flattening of the shape undulations.
This leads to an increase in the
the membrane tension \cite{Evans-Rawicz:1990}
\begin{equation}
\label{entropic tens}
 \sigma_0= \sigma_{in}\exp\left[\frac{8 \pi \kappa }{k_B T}\left(\frac{\bar A(t)}{4 \pi a ^2}-1\right)\right]\,,
 \end{equation}
where $\sigma_{in}$ is the initial membrane tension.

In the next section we simplify the theory for the case of vesicle electrodeformation induced by
an uniform AC electric field.

\section{Results}
\label{sec3}

\subsection{Deformation of a quasi-spherical vesicle}

When the electric field is turned on,  it generates electrohydrodynamic flow with the same symmetry as the electric stresses. The corresponding fluid velocity, which is responsible for the vesicle deformation, is given by
\begin{equation}
C^\el\equiv C_{20}=8\sqrt{\frac{\pi}{5}}\frac{6\tau_r^\el-\tau_\theta^\el}{23 \visrat+16\visrat_s+32}\,,
\end{equation}
where the electric stresses are given by \refeq{tn} and  \refeq{tt} in Appendix \ref{el field shell}, and
the viscosity parameters $\visrat$ and  $\visrat_s$ are defined by \refeq{visrat} and \refeq{visratS}.
Since electric stresses directly affect only  the ellipsoidal $j=2,\, m=0$ mode,
the most important contribution to the vesicle deformation comes from the ``elongational'' $f_{20}$ mode. Moreover, because the shape  modes are coupled through the area constraint \refeq{excess area},
the  area stored in the $j\neq 2$ modes is transferred into the ellipsoidal $f_{20}$ mode.
The maximum possible vesicle deformation corresponds to elongation where all excess area is stored in the $f_{20}$ mode
\begin{equation}
\label{steady def}
f_{20}^{max}=\pm \sqrt{\frac{\Delta}{2}}\,,
\end{equation}
where a positive sign corresponds to a prolate deformation.

The shape evolution strongly depends on the effective tension $\sigma_0$.
For a quasi-spherical vesicle,
using the relation between excess area and shape modes \refeq{excess area}, and including only the dominant contribution from the $f_{20}$ mode, we can rewrite \refeq{entropic tens} as
\begin{equation}
\label{entropic tens:2}
 \sigma_0= \sigma_{in}\exp\left(\frac{4 \kappa }{k_B T}f_{20}^2\right)\,.
 \end{equation}
Inserting into \refeq{ev equation f} we obtain  that the shape evolution of a vesicle in AC electric field is described by the following non-linear equation
\begin{equation}
\label{f20AC-1}
\frac{\partial f_{20}}{\partial t}=C^\el-\Ca^{-1} \frac{24\left[6+\sigma_{in}\exp\left(\frac{4 \kappa }{k_B T}f_{20}^2(t)\right)\right]}{23\visrat+16\visrat_s+32}f_{20}(t)\,.
\end{equation}
Our theory can also be applied to  vesicles with  non-spherical rest shapes, i.e., non-zero initial apparent area,  as shown in  Appendix~\ref{strong fields}.

\subsection{Discussion}
\label{Discussion}

The shape evolution obtained from \refeq{f20AC-1} for several frequencies is illustrated in  Figure \ref{shapef20}.a.
The vesicle deforms on a hydrodynamic time scale approximately given  by  $t_d=1/C^\el$.
The time needed to reach stationary shape depends strongly on the viscosity contrast between the inner and outer fluid.
Figure~\ref{shapef20}.b  shows that increasing the viscosity of the inner fluid slows down the shape evolution.
The viscosity effect  may become important in the electrodeformation of red blood cells, which are characterized by $\visrat\sim10$. Another factor that can slow down the shape evolution even more dramatically is  the membrane viscosity, as illustrated in Figure~\ref{shapef20}.c. In the case of polymersomes or lipid membranes undergoing fluid-to-gel transition, the membrane viscosity parameter can reach values of the order of 100 \cite{Dimova:2000}.  The sensitivity of shape evolution to membrane viscosity suggests a novel method for determination of the membrane viscosity  where the experimental effort is minimal.

The steady shape of a vesicle in AC electric field is calculated by evaluating \refeq{f20AC-1}.
Figure \ref{shapesAC} illustrates the steady shapes of vesicles in AC field as a function of frequency for
different conductivity ratios.  The theory predicts that the type of deformation, prolate or oblate, is determined primarily by the frequency and the conductivity ratio.
At low frequencies $\omega<\omega_1$ the deformation is prolate. For frequencies $\omega>\omega_1$ vesicles are prolate or oblate depending on the conductivity ratio. At even higher frequencies,
the deformation becomes again prolate but very small and the vesicle appears spherical. This observation corresponds
well to the experimental data.

Next we analyze these morphological transitions in more detail.

\subsubsection{Prolate-oblate transition for $\Lambda<1$ at low frequencies}

The transition frequency $\omega_1$ corresponds to the capacitor charging time \refeq{omega1}.

At low frequencies, $\omega<\omega_1$, vesicle deformation is due solely to the positive electric pressure. It is maximal at the poles, see \refeq{tn} in Appendix \ref{el field shell}. The vesicle is pulled apart and thus adopts a prolate ellipsoidal shape.

At  $\omega>\omega_1$, the tangential electric traction becomes significant and the electric pressure is negative,  as seen from Figure \ref{fig:tractions}. The shearing tangential force induces electrohydrodynamic flow, similar to the one observed with drops (Figure \ref{fig: toroid}).
If $\Lambda/S<1$ the  flow is directed  from the poles to the  equator and the resulting deformation is oblate; if  $\Lambda/S>1$ the  flow is directed  from the  equator to  the poles and the resulting deformation is prolate.  Therefore, prolate-oblate transition is possible only if $\Lambda/S<1$.  In experiments with vesicles\cite{Riske-Dimova:2006, Dimova-Aranda:2007, Aranda:2007}, the inner and outer fluids are sucrose and glucose, which have similar dielectric constant, $S\sim 1$. Oblate shapes were reported for conductivity ratio less than 1, in agreement with the condition $\Lambda/S<1$. In the case of biological cells, the difference between the dielectric constants of the cytosol and the cell environment is also small, and therefore similar deformation behavior is expected.
In the case of drops, the electrohydrodynamic flow persists for as long as the electric field is applied because only viscous stresses can balance the tangential  electric surface force.  In contrast to drops,  the electrohydrodynamic flow in vesicles is not sustained. It stops
when the vesicle reaches
steady deformation because the  membrane tension counteracts the electric tangential force.

The capacitor charging time decreases with the size of the vesicle. Therefore, the smaller the vesicle, the higher the transition frequency. For nanometer size vesicle this frequency is in the MHz range. Thus, nano-vesicles are expected to deform only into prolate ellipsoids when subjected to AC fields with frequency less than a MHz or DC pulses with length longer than 1 $\mu s$, which is in agreement with experimental observations \cite{Kakorin:2003}.
The theoretical predictions for the prolate-oblate transition frequency observed for giant vesicles are in good agreement with  experiments \cite{Aranda:2007}, as shown in Figure \ref{fig: trans freq}. Note that the reported experimental data was collected for vesicles with various sizes and conductivity conditions.

\subsubsection{Oblate-prolate transition for $\Lambda<1$ at high frequencies }
\label{oblprol}

At high frequencies $\omega \gg \omega_1$, the transmembrane potential vanishes, as seen in  Figure \ref{fig:potential}.
The electric tractions are given by the zero-thickness results \refeq{tn0} and \refeq{tt0} in Appendix \ref{el field shell}.
The forcing term $C^{\el}$ in the shape evolution \refeq{f20AC-1} changes sign at a frequency
\begin{equation}
\label{ehdfreq}
\Omega_2=\left[\frac{4S-(\Lambda+1)^2}{(S-1)^2}\right]^\half\,.
\end{equation}
Correspondingly, the vesicle deformation changes from oblate to prolate at this frequency.
The transition frequency $\Omega_2$ becomes very large when the dielectric constants of  the fluids are comparable. For vesicles filled with sucrose and suspended in glucose solutions this frequency is about 10MHz, which is in the frequency range where electric tractions have already become too small to deform the vesicle.  Thus this oblate-prolate transitions was not observed in the experiments of Aranda et al. \cite{Aranda:2007}; instead, the vesicles remain spherical. Thus far, the prolate-oblate transition has been reported only for drops \cite{Torza}.

If $\Omega_2<\omega_1$, the oblate deformation would be impossible. This situation arises if the membrane becomes highly conducting, e.g., because of poration. Another possibility is a thick membrane or small vesicle with
\begin{equation}
\frac{h}{a}>\frac{(\Lambda+2)^2S_m} {(S+2)(2\Lambda+G_m(\Lambda+2))}
\end{equation}
where  $S_m=\epsilon_\mem/\epsilon_\out$. For a typical bilayer thickness of $5nm$ this condition holds for  vesicle size below $100nm$. This prediction is in agreement with experimental studies of nano-sized vesicles \cite{Kakorin:2003} that have reported only prolate deformations.

The oblate--prolate transition is independent of membrane properties;
it is analogous to the one observed with drops \cite{Torza, Vizika:1992}.
It is also independent of the viscosity ratio because the electrohydrodynamic flow stops at steady state due the interface
immobilization  by gradients in the membrane tension.

\subsubsection{The effective dipole theory does not predict the prolate-oblate transition}

The effective dipole theory, summarized in Appendix \ref{eff dipole},
models the cell as a sphere with effective permittivity.
The theory  successfully  explains the dielectrophoresis and electrorotation of cells, because it correctly describes the perturbation due to the cell in the exterior electric field. However, the  internal electric field is not physical, which leads  to incorrect interior Maxwell stress and electric force distribution on the membrane.
Accordingly, the predicted deformation is  oblate at low frequencies \cite{Sukhorukov:1998}, which is at odds with the experimental observations with vesicles \cite{Aranda:2007}.

Figure \ref{dipole_comp} compares the predictions of our model and the effective dipole theory for the electric  tractions. It shows that  the two models agree at frequencies $\omega>\omega_1$, where the transmembrane potential has vanished.
At low frequencies, where the field inside the vesicle is zero, the effective dipole theory  would correctly predict  the electric tractions  if only the contribution from the exterior electric field is taken into account. However, at intermediate frequencies,  where the vesicle interior participates in the conduction process and the transmembrane potential is still significant, i.e., $\omega \sim \omega_1 $, the effective dipole theory diverges from our model as well as experimental observations~\cite{Aranda:2007}.

\section{Conclusions and outlook}

We have developed a theory that explains the observed morphological transitions of vesicles in a uniform AC electric field, in particular, the shape dependence  on
the field frequency and conductivity ratio between the inner and outer fluids. Prolate deformations at low  frequencies
have purely dielectric origin and result from electric pressure due to polarization charges pulling the vesicle at the poles. Oblate deformations, however,
result from induced  free surface charges, which cause negative pressure
 and transient electrohydrodynamic flow driven by tangential electric tractions.
The  prolate-oblate transition at low frequencies depends on the membrane capacitance and conductance.
At high frequencies, electric stresses become negligible and do not affect the vesicle equilibrium quasi-spherical shape.
The theory also predicts  a  high-frequency  oblate-prolate transition, which is analogous to the one observed with drops: it is independent of the membrane electric properties and depends only on the conductivities of inner and outer fluids. The transition frequency, however, is not given by the Maxwell-Wagner polarization time,
but  is determined by electrohydrodynamics.

We have considered the problem for vesicle electrodeformation from a mechanical point of view where the vesicle shape is determined by the balance of  forces exerted on the interface.
Thus,  our formalism  can be easily extended to electric fields of arbitrary symmetry as well as to situations when external electric and flow fields are simultaneously applied.

Our current  theory is a step in a systematic study of the electrohydrodynamics of deformable cells and, as such,  some potentially important effects are neglected.
First, our treatment assumes that all media are electrically homogeneous and is based on solutions of Laplace's equation. This approach requires that the Debye length of the media is small compared to the radius of the vesicle or the thickness of the membrane.  Thus our  theory might break down  at low conductivities and frequencies.  Second, the model does not include  shear elasticity of the membrane, which is essential in the mechanics of the red blood cell.
Third, the  membrane is assumed to be non-permeable to ions. However, at low frequencies,  the duration of application of the electric field  may be sufficient to porate the membrane.
An electric current due to ion movement through field-induced pores  would affect the electric field and tractions, and therefore vesicle shapes.
Electrokinetic effects, the role of shear elasticity, membrane poration, and membrane charge represent interesting and challenging problems to be investigated in the future.

\section{Acknowledgments}

PV thanks Thomas Powers and Margarita Staykova for stimulating discussions.

\appendix
\section{List of symbols}
subscript $r$ denotes {\it{radial}}\\
subscript $\theta$ denotes {\it{tangential}}\\
sub/superscript ``$\el$'' denotes {\it{electric}}\\
sub/superscript ``$\hd$'' denotes {\it{hydrodynamic}}\\
sub/superscript ``$\mem$'' denotes {\it{membrane}}\\
sub/superscript ``$\ins$'' denotes {\it{interior}}\\
sub/superscript ``$\out$'' denotes {\it{exterior}}\\
superscript $*$ denotes {\it{complex conjugate}}\\
$Re[ ]$ denotes Real part of [ ]\\
$Im[ ]$ denotes Imaginary part of [ ]\\
\\
$a$ vesicle radius\\
$C_m$ membrane capacitance\\
$\Ca$ capillary number\\
$E$ electric field\\
$f_{jm}$ shape deformation parameter\\
$G_m$ membrane conductivity \\
$h$ membrane thickness\\
$H$ mean curvature\\
$p$ pressure\\
$S$ permittivity ratio\\
$t$ time\\
$\bT$ bulk stress\\
$\bv$ fluid velocity\\
$V_m$ transmembrane potential\\
$x=h/a$ dimensionless membrane thickness\\
$Y_{jm}$ spherical harmonic\\
\\
$\eta$  viscosity\\
$\rho$ density
$\lambda$  conductivity\\
$\epsilon$  permittivity\\
$\Lambda$ conductivity ratio\\
$\chi$ viscosity ratio\\
$\chi_s$ membrane viscosity parameter\\
$\omega_1$ frequency of the prolate-oblate transition\\
$\omega_2$ frequency corresponding to the inverse Maxwell-Wagner polarization time\\
$\Omega_2$ frequency of the oblate-prolate transition\\
$\Phi$ electric potential\\
$\tau$  tractions\\
$\sigma$ membrane tension\\
$\kappa$ bending rigidity\\
$\Delta$ excess area\\

\section{Spherical harmonics}
\label{Harmonics}

The normalized spherical scalar harmonics are defined as \cite{Varshalovich:1988}
\begin{equation}
\label{normalized spherical harmonics}
   Y_{jm}\brac{\theta,\varphi} = \textstyle \left[\frac{2j+1}{4\pi}\frac{(j-m)!}{(j+m)!}\right] (-1)^m P_j^m(\cos\theta)e^{{\rm i}m\varphi},
\end{equation}
where   $(r, \theta,\varphi)$
are the spherical coordinates, and $P_j^m(\cos\theta)$ are the Legendre polynomials.
For example
\begin{equation}
Y_{10}=\frac{1}{\sqrt{4 \pi}}\cos\theta \,.
\end{equation}

\section{Electrostatic field and stresses for a spherical shell}

\subsection{Our model: A sphere with interfacial capacitance and conductivity}
\label{el field shell}
Schwann et al. \cite{Grosse-Schwan:1992, Schwan} have solved the problem for the electric field about a spherical shell with radius $a$ and shell thickness $h$ to obtain \refeq{potential} for the potential difference between the inner and outer shell surfaces. Assuming a very thin shell $h/a\ll1$, we can approximate the membrane with a two-dimensional interface that possesses capacitance. Accordingly, the spherical shell is approximated by a sphere with a discontinuous potential at the interface.

Solving \refeq{Maxwell eqs} with the boundary conditions \refeq{potential}, \refeq{normal balance} and \refeq{charge conservation:AC} leads to
\begin{equation}
\label{el field}
\Phi^\out=-[r + P^\out r^{-2}]\exp(\im \omega t)\cos\theta\,,\qquad
\Phi^\ins=-P^\ins r \exp(\im \omega t)\cos\theta\,.
\end{equation}
where
\begin{equation}
\label{el field2}
P^\out= d^\infty\frac{(-k_\ins+k_\out)+k_\ins V_m}{k_\ins+2k_\out}\,,\qquad
P^\ins= d^\infty k_\out\frac{3-2V_m}{k_\ins+2k_\out}\,,
\end{equation}
and $k$ denote the dimensionless complex conductivities of the inner and outer fluids
\begin{equation}
k_\ins={\Lambda+\im \omega S }\,,\quad k_\out={1+\im \omega} \,.
\end{equation}

The tractions are computed from the Maxwell stress tensor.
The radial (pressure) component is given by
\begin{equation}
\label{tn}
\tau_r^\el=\textstyle\frac{1}{32\pi} [-2(\tau_1^2+\tau_2^2)S
+5 \tau_3^2-2 d^\infty \tau_3+5 \tau_4^2+2(d^\infty)^2]\,,
\end{equation}
and the tangential (shearing) component is
\begin{equation}
\label{tt}
\tau_\theta^\el=\textstyle -\frac{3}{8\pi} \left[(\tau_1^2+\tau_2^2)S+2 \tau_3^2+(d^\infty)\tau_3+2 \tau_4^2-(d^\infty)^2\right] \,,
\end{equation}
where $\tau_1=Re[P^\ins]\,,\tau_2=Im[P^\ins], \tau_3=Re[P^\out]\,,\tau_4=Im[P^\out] $. $Re[\,]$ and $Im[\,]$ denote real and imaginary part.
Taking the zero-thickness limit,  $x=0$, our solution reduces to  the result for a spherical drop   \cite{Torza}
\begin{equation}
\label{tn0}
\tau_r^{\el, drop}=\textstyle   \frac{3}{8} \left(1+ \Lambda^2-2S+(S-1)^2 S \omega^2\right) (2 +S)^2(\omega^2+\omega_2^2)^{-1}\,,
\end{equation}
\begin{equation}
\label{tt0}
\tau_\theta^{\el, drop}=\textstyle \frac{9}{2}(\Lambda-S)(2+S)^2(\omega^2+\omega_2^2)^{-1}\,,
\end{equation}
where $\omega_2$ is given by \refeq{omega2}

The effective charge density is calculated from \refeq{normal balance}
\begin{equation}
Q(\omega,t)=q_c(\omega) \cos(\omega t)+q_s(\omega)\sin(\omega t) \,,
\end{equation}
where
\begin{equation}
q_s(\omega)=2\tau_4+S \tau_2\,,\qquad q_s(\omega)=d^\infty-2\tau_3-S\tau_1
\end{equation}
The frequency dependence of the charge density can be cast into the form
\begin{equation}
Q(\omega,t)=\bar Q(\omega)\cos(\omega t+\psi) \,,
\end{equation}
where the amplitude is $\bar Q(\omega)=[q_s+q_c]^{\half}$, and the phase shift is $\psi=q_s/q_c$.

\subsection{The effective sphere model}
\label{eff dipole}

The  dipole theory
models the cell as a sphere with an
 effective permittivity \cite{Jones:2003, Schwan:1992}
\begin{equation}
k_\ins^{eff}=k_\ins\left[(1-x)^{-3}+2\frac{k_\ins-k_\mem}{k_\ins+2k_\mem}\right]\left[(1-x)^{-3}-\frac{k_\ins-k_\mem}{k_\ins+2k_\mem}\right]^{-1}\,,
\end{equation}
where
\begin{equation}
k_\ins={\Lambda+\im \omega S }\,,\quad k_\mem={\Lambda_\mem+\im \omega S_\mem} \,.
\end{equation}
The  electric field is described by an electric potential
\begin{equation}
\Phi^\out= -d^\infty\left[r + r^{-2}\frac{(-k^{eff}_\ins+k_\out)}{k^{eff}_\ins+2k_\out}\right]\,,\qquad
\Phi^\ins= - r d^\infty k_\out\frac{3}{k^{eff}_\ins+2k_\out}\cos\theta\,,
\end{equation}
The electric potential is continuous, and hence there is no transmembrane potential within the framework of the effective dipole theory.


\section{Deformation of a prolate vesicle in strong fields}
\label{strong fields}


Consider an initially non-spherical, non-fluctuating vesicle. This situation can occur in strong electric fields, where the vesicle is already maximally deformed, \refeq{steady def}, and then the field direction is changed. The evolution to the new stationary shape is no longer described by \refeq{f20AC-1} because the tension is no longer given by \refeq{entropic tens}. The effective tension  has to be determined self-consistently along with the field-induced changes in shape to keep the total area constant \cite{Vlahovska:2007}, see Supplementary material. The leading order vesicle electrohydrodynamics becomes non--linear in contrast to the corresponding results for drops and capsules \cite{Taylor:1966, Torza, Ha:2000c}. This feature of non-equilibrium vesicle dynamics has been noted by several authors in relation to vesicle dynamics in shear flow \cite{Misbah:2006, Vlahovska:2007, Lebedev:2008}.

The vesicle deformation described  by \refeq{ev equation f} can  be approximated by
\begin{equation}
\label{f2m evolution}
\dot f_{20}=C^\el(1-2 \Delta^{-1}f^2_{20})\,\qquad
\dot f_{2m}=-2C^\el \Delta^{-1}f_{20}f_{2m}
\end{equation}
where the dot denotes time derivative. The modes $f_{2m}$ are slaved to the $f_{20}$,
which is forced to change by the electric field.
\refeq{f2m evolution} can be integrated to yield
\begin{equation}
\label{f20AC}
f_{20}(t)=\delta\tanh\left[\frac{ C^\el}{\delta} t+\tanh^{-1}\left(\frac{f_{20}(0)}{\delta}\right)\right].
\end{equation}
where $\delta$ is the maximum possible deformation
\begin{equation}
\label{delta}
\delta=\sqrt{\frac{\Delta}{2}}\,.
\end{equation}.





\clearpage

\bibliography{refs}

\begin{thebibliography}{63}
\providecommand{\url}[1]{\texttt{#1}}
\providecommand{\urlprefix}{ }

\bibitem[Aranda et~al.(2008)Aranda, Riske, Lipowsky, and Dimova]{Aranda:2007}
Aranda, S., K.~A. Riske, R.~Lipowsky, and R.~Dimova, 2008.
\newblock Morphological transitions of vesicles induced by AC electric fields.
\newblock \emph{Biophys. J.} 95:L19--L21.

\bibitem[Zhao et~al.(2006)Zhao, Song, Pu, Wada, Reid, Tai, Wang, Guo,
  Walczysko, Gu, Sasaki, Suzuki, Forrester, Bourne, Devreotes, McCaig, and
  Penninger]{Zhaoetal}
Zhao, M., B.~Song, J.~Pu, T.~Wada, B.~Reid, G.~P. Tai, F.~Wang, A.~H. Guo,
  P.~Walczysko, Y.~Gu, T.~Sasaki, A.~Suzuki, J.~V. Forrester, H.~R. Bourne,
  P.~N. Devreotes, C.~D. McCaig, and J.~M. Penninger, 2006.
\newblock Electrical signals control wound healing through
  phosphatidylinositol-3-OH kinase- and PTEN.
\newblock \emph{Nature} 442:457--460.

\bibitem[Funk and Monsees(2006)]{Funk}
Funk, R. H.~W., and T.~K. Monsees, 2006.
\newblock Effects of Electromagnetic Fields on Cells: Physiological and
  Therapeutical Approaches and Molecular Mechanisms of Interaction.
\newblock \emph{Cells Tissues Organs} 182:59--78.

\bibitem[Voldman(2006)]{Voldman:2006}
Voldman, J., 2006.
\newblock Electrical Forces For Microscale Cell Manipulation.
\newblock \emph{Annu. Rev. Biomed. Eng.} 8:425--454.

\bibitem[Zimmermann and Neil(1996)]{Zimmermann-Neil:1996}
Zimmermann, U., and G.~A. Neil, 1996.
\newblock Electromanipulation of cells.
\newblock CRC Press, Boca Raton.

\bibitem[Neumann et~al.(1989)Neumann, Sowers, and
  Jordan]{Neumann-Sowers-Jordan:1989}
Neumann, E., A.~E. Sowers, and C.~A. Jordan, 1989.
\newblock Electroporation and electrofusion in cell biology.
\newblock Plenum Press, New York.

\bibitem[Chizmadzhev et~al.(1985)Chizmadzhev, Kuzmin, and
  Pastushenko]{Chizmadzhev-Kuzmin:1985}
Chizmadzhev, Y.~A., P.~Kuzmin, and V.~P. Pastushenko, 1985.
\newblock Theory of the dielectrophoresis of vescicles and cells.
\newblock \emph{Biol. Mem.} 2:1147--1161.

\bibitem[Chizmadzhev et~al.(1988)Chizmadzhev, Kuzmin, and
  Pastushenko]{Pastushenko:1988}
Chizmadzhev, Y.~A., P.~Kuzmin, and V.~P. Pastushenko, 1988.
\newblock Dielectrophoresis and electrorotation of cells: unified theory for
  spherically symmetric cells with arbitrary structure of membrane.
\newblock \emph{Biol. Mem.} 5:65--78.

\bibitem[Turcu and Lucaciu(1989)]{Turcu}
Turcu, I., and C.~M. Lucaciu, 1989.
\newblock Dielectrophosresis -a spherical shell model.
\newblock \emph{J. Phys. A} 22:985--993.

\bibitem[Jones(1995)]{JonesTB}
Jones, T.~B., 1995.
\newblock Electromechanics of particles.
\newblock Cambridge University Press, New York.

\bibitem[Gimsa and Wachner(1999)]{Gimsa-Wachner:1999}
Gimsa, J., and D.~Wachner, 1999.
\newblock A polarization model overcoming the geometric restrictions of the
  laplace solution for spheroidal cells: Obtaining new equations for
  field-induced forces and transmembrane potential.
\newblock \emph{Biophys. J.} 77:1316--1326.

\bibitem[Dolinsky and Elperin(2006)]{Dolinsky-Elperin:2006}
Dolinsky, Y., and T.~Elperin, 2006.
\newblock Dynamics of a spheroidal particle in a leaky dielectric medium in an
  ac electric field.
\newblock \emph{Phys. Rev. E} 73:066607.

\bibitem[Foster et~al.(1992)Foster, Sauer, and Schwan]{Schwan:1992}
Foster, K.~R., F.~A. Sauer, and H.~P. Schwan, 1992.
\newblock Electrorotation and levitation of cells and colloidal particles.
\newblock \emph{Biophys. J.} 63:180--190.

\bibitem[Jones(2003)]{Jones:2003}
Jones, T.~B., 2003.
\newblock Basic theory of dielectrophoresis and electrorotation.
\newblock \emph{IEEE Eng Med Biol Mag.} 22:33--42.

\bibitem[Miller and Jones(1993)]{Miller-Jones:1993}
Miller, R.~D., and T.~B. Jones, 1993.
\newblock Electro-orientation of ellipsoidal erythrocytes. Theory and
  experiment.
\newblock \emph{Biophys. J.} 64:1588--1595.

\bibitem[Engelhardt and Sackmann(1988)]{Engelhardt-Sackmann:1988}
Engelhardt, H., and E.~Sackmann, 1988.
\newblock On the measurement of shear elastic moduli and viscosities of
  erythrocyte plasma membranes by transient deformation in high frequency
  electric fields.
\newblock \emph{Biophys J.} 54:495--508.

\bibitem[Bryant and Wolfe(1987)]{Bryant-Wolfe:1987}
Bryant, G., and J.~Wolfe, 1987.
\newblock Electromechanical stresses produced in the plasma membranes of
  suspended cells by applied electric fields.
\newblock \emph{J. Membr. Biol.} 96:129--139.

\bibitem[Poznanski et~al.(1992)Poznanski, Pawlowski, and Fikus]{Pawlowski:1992}
Poznanski, J., P.~Pawlowski, and M.~Fikus, 1992.
\newblock Bioelectrorheological model of the cell. 3. Viscoelastic shear
  deformation of the membrane.
\newblock \emph{Biophys. J.} 61:612--620.

\bibitem[Helfrich(1974)]{Helfrich:1974}
Helfrich, W., 1974.
\newblock Deformation of lipid bilayer spheres by electric fields.
\newblock \emph{Z. Naturforsch.} 29c:182--183.

\bibitem[Kummrow and Helfrich(1991)]{Kummrow-Helfrich:1991}
Kummrow, M., and W.~Helfrich, 1991.
\newblock Deformation of giant lipid vesicles by electric fields.
\newblock \emph{Phys. Rev. A} 44:8356--8360.

\bibitem[Winterhalter and Helfrich(1988)]{Winterhalter-Helfrich:1988}
Winterhalter, M., and W.~Helfrich, 1988.
\newblock Deformation of spherical vesicles by electric fields.
\newblock \emph{J. Coll. Int. Sci.} 122:583--586.

\bibitem[Teissie and Tsong(1981)]{Teissie:1981}
Teissie, J., and T.~Y. Tsong, 1981.
\newblock Electric field induced transient pores in phospholipid bilayer
  vesicles.
\newblock \emph{Biochemistry} 20:1548--1554.

\bibitem[Sukhorukov et~al.(1998)Sukhorukov, Mussauer, and
  Zimmermann]{Sukhorukov:1998}
Sukhorukov, V., H.~Mussauer, and U.~Zimmermann, 1998.
\newblock The effect of electrical deformation forces on the
  electropermeabilization of erythrocyte membranes in low- and
  high-conductivity media.
\newblock \emph{J. Memb. Biol.} 163:235--245.

\bibitem[Isambert(1998)]{Isambert:1998}
Isambert, H., 1998.
\newblock Understanding the electroporation of cells and artificial bilayer
  membranes.
\newblock \emph{Phys. Rev. Lett.} 80:3404--3407.

\bibitem[Sens and Isambert(2002)]{Sens-Isambert:2002}
Sens, P., and H.~Isambert, 2002.
\newblock Undulation instability of lipid membranes under an electric field.
\newblock \emph{Phys. Rev. Lett.} 88:Art. No. 128102.

\bibitem[Lacoste et~al.(2007)Lacoste, Lagomarsino, and Joanny]{Lacoste:2007}
Lacoste, D., M.~Lagomarsino, and J.~Joanny, 2007.
\newblock Fluctuations of a driven membrane in an electrolyte.
\newblock \emph{Europhys. Lett.} 77:18006.

\bibitem[Ambjornsson et~al.(2007)Ambjornsson, Lomholt, and
  Hansen]{Ambjornsson:2007}
Ambjornsson, T., M.~A. Lomholt, and P.~L. Hansen, 2007.
\newblock Applying a potential across a biomembrane: Electrostatic contribution
  to the bending rigidity and membrane instability.
\newblock \emph{Phys. Rev. E} 75:051916.

\bibitem[Krassowska and Filev(2007)]{Krassowska:2007}
Krassowska, W., and P.~Filev, 2007.
\newblock Modeling electroporation in a single cell.
\newblock \emph{Biophys. J.} 92:404--417.

\bibitem[Riske and Dimova(2005)]{Riske-Dimova:2005}
Riske, K.~A., and R.~Dimova, 2005.
\newblock Electro-deformation and poration of giant vesicles viewed with high
  temporal resolution.
\newblock \emph{Biophys. J.} 88:1143--1155.

\bibitem[Riske and Dimova(2006)]{Riske-Dimova:2006}
Riske, K.~A., and R.~Dimova, 2006.
\newblock Electric pulses induce cylindrical deformations on giant vesicles in
  salt solutions.
\newblock \emph{Biophys. J.} 91:1778--1786.

\bibitem[Dimova et~al.(2006)Dimova, Aranda, Bezlyepkina, Nikolov, Riske, and
  Lipowsky]{Rumy:2006}
Dimova, R., S.~Aranda, N.~Bezlyepkina, V.~Nikolov, K.~A. Riske, and
  R.~Lipowsky, 2006.
\newblock A practical guide to giant vesicles. Probing the membrane nanoregime
  via optical microscopy.
\newblock \emph{J. Phys. Cond. Matt.} 18:S1151--S1176.

\bibitem[Dimova et~al.(2007)Dimova, Riske, Aranda, Bezlyepkina, Knorr, and
  Lipowsky]{Dimova-Aranda:2007}
Dimova, R., K.~A. Riske, S.~Aranda, N.~Bezlyepkina, R.~L. Knorr, and
  R.~Lipowsky, 2007.
\newblock Giant vesicles in electric fields.
\newblock \emph{Soft matter} 3:817--827.

\bibitem[Mitov et~al.(1993)Mitov, Meleard, Winterhalter, Angelova, and
  Bothorel]{Mitov:1993}
Mitov, M.~D., P.~Meleard, M.~Winterhalter, M.~I. Angelova, and P.~Bothorel,
  1993.
\newblock Electric-field-dependent thermal fluctuations of giant vesicles.
\newblock \emph{Phys.\ Rev.\ E} 48:628--631.

\bibitem[Peterlin et~al.(2007)Peterlin, Svetina, and Zeks]{Peterlin:2007}
Peterlin, P., S.~Svetina, and B.~Zeks, 2007.
\newblock The prolate-to-oblate shape transition of phospholipid vesicles in
  response to frequency variation of an AC electric field can be explained by
  the dielectric anisotropy of a phospholipid bilayer.
\newblock \emph{J. Phys. Cond. Phys.} 19:136220.

\bibitem[Hyuga et~al.(1991)Hyuga, Kinosita~Jr., and Wakabayashi]{Hyuga:1991a}
Hyuga, H., K.~Kinosita~Jr., and N.~Wakabayashi, 1991.
\newblock Deformation of vesicles under the influence of strong electric
  fields.
\newblock \emph{Jpn. J. Appl.Phys.} 30:1141--1148.

\bibitem[Nelson(2004)]{Nelson}
Nelson, P., 2004.
\newblock Life in the Slow Lane: The Low Reynolds-Number World.
\newblock \emph{In} Biological Physics: Energy, Information, Life, Freeman,
  158--194.

\bibitem[Sauer(1985)]{Sauer}
Sauer, F.~A., 1985.
\newblock Interaction forces between microscopic particles in an external
  electromagnetic field.
\newblock \emph{In} A.~Chiabrera, C.~Nicolini, and H.~P. Schwan, editors,
  Interactions between electromagnetic fields and cells, Plenum Press,
  181--202.

\bibitem[Taylor(1966)]{Taylor:1966}
Taylor, G.~I., 1966.
\newblock Studies in electrohydrodynamics. I. Circulation produced in a drop by
  an electric field.
\newblock \emph{Proc. Royal Soc. A} 291:159--166.

\bibitem[Melcher and Taylor(1969)]{Melcher-Taylor:1969}
Melcher, J.~R., and G.~I. Taylor, 1969.
\newblock Electrohydrodynamics - a review of role of interfacial shear stress.
\newblock \emph{Annu. Rev. Fluid Mech.} 1:111--146.

\bibitem[Saville(1997)]{Saville:1997}
Saville, D.~A., 1997.
\newblock Electrohydrodynamics: The Taylor-Melcher leaky dielectric model.
\newblock \emph{Annu. Rev.Fluid Mech.} 29:27--64.

\bibitem[Schwan(1989)]{Schwan}
Schwan, H.~P., 1989.
\newblock Dielectrophoresis and rotation of cells.
\newblock \emph{In} E.~Neumann, A.~E. Sowers, and C.~A. Jordan, editors,
  Electroporation and electrofusion in cell biology, Plenum Press, 3--21.

\bibitem[Kinosita~Jr. et~al.(1988)Kinosita~Jr., Ashikawa, Saita, Yoshimura,
  Itoh, Nagayama, and Ikegami]{Kinosita:1988}
Kinosita~Jr., K., I.~Ashikawa, N.~Saita, H.~Yoshimura, H.~Itoh, K.~Nagayama,
  and A.~Ikegami, 1988.
\newblock Electroporation of cell membrane visualized under a pulsed laser
  fluorescence microscope.
\newblock \emph{Biophys. J.} 53:1015--1019.

\bibitem[Needham and Hochmuth(1989)]{Needham-Hochmuth:1989}
Needham, D., and R.~M. Hochmuth, 1989.
\newblock Electromechanical permeabilization of lipid vesicles. Role of
  membrane tension and compressibility.
\newblock \emph{Biophys. J.} 55:1001--1009.

\bibitem[Leal(1980)]{Leal:1980}
Leal, L.~G., 1980.
\newblock Particle motions in a viscous fluid.
\newblock \emph{Ann. Rev. Fluid Mech} 12:435--476.

\bibitem[Sozou(1972)]{Sozou:1972}
Sozou, C., 1972.
\newblock Electrohydrodynamics of a liquid drop - time-dependent problem.
\newblock \emph{Proc. Royal Soc. A} 331:263--272.

\bibitem[Leal(1992)]{Leal:1992}
Leal, L.~G., 1992.
\newblock Laminar Flow and Convective Transport Processes.
\newblock Butterworth-Heinemann, Boston.

\bibitem[Petrov(2006)]{Petrov:review}
Petrov, A.~G., 2006.
\newblock Electricity and mechanics of biomembrane systems: Flexoelectricity in
  living membranes.
\newblock \emph{Analytica Chimica Acta} 568:70--83.

\bibitem[Raphael et~al.(2000)Raphael, Popel, and Brownell]{Raphael:2000}
Raphael, R.~M., A.~S. Popel, and W.~E. Brownell, 2000.
\newblock A Membrane Bending Model of Outer Hair Cell Electromotility.
\newblock \emph{Biophys. J.} 78:2844--–2862.

\bibitem[Seifert(1999)]{Seifert:1999}
Seifert, U., 1999.
\newblock Fluid membranes in hydrodynamic flow fields: Formalism and an
  application to fluctuating quasispherical vesicles.
\newblock \emph{Eur. Phys. J. B} 8:405--415.

\bibitem[Seifert(1997)]{Seifert:1997}
Seifert, U., 1997.
\newblock Configurations of fluid membranes and vesicles.
\newblock \emph{Advances in physics} 46:13--137.

\bibitem[Discher and Ahmed(2006)]{Discher:2006}
Discher, D.~E., and F.~Ahmed, 2006.
\newblock Polymersomes.
\newblock \emph{Annu. Rev. Biomed. Eng.} 8:323--341.

\bibitem[Dimova et~al.(2002)Dimova, Seifert, Poligny, Forster, and
  Dobereiner]{Rumy:2002}
Dimova, R., U.~Seifert, B.~Poligny, S.~Forster, and H.-G. Dobereiner, 2002.
\newblock Hyperviscous diblock coploymer vesicles.
\newblock \emph{Eur. Phys. J. D} 7:241--250.

\bibitem[Grosse and Schwan(1992)]{Grosse-Schwan:1992}
Grosse, C., and H.~P. Schwan, 1992.
\newblock Cellular membrane potentials induced by alternating fields.
\newblock \emph{Biophys. J.} 63:1632--1642.

\bibitem[Vlahovska and Gracia(2007)]{Vlahovska:2007}
Vlahovska, P.~M., and R.~Gracia, 2007.
\newblock Dynamics of a viscous vesicle in linear flows.
\newblock \emph{Phys. Rev. E} 75:016313.

\bibitem[Evans and Rawicz(1990)]{Evans-Rawicz:1990}
Evans, E., and W.~Rawicz, 1990.
\newblock Entropy driven tension and bending elasticity in condensed-fluid
  membranes.
\newblock \emph{Phys.\ Rev.\ Lett.} 64:2094--2097.

\bibitem[Dimova et~al.(2000)Dimova, Poligny, and Dietrich]{Dimova:2000}
Dimova, R., B.~Poligny, and C.~Dietrich, 2000.
\newblock Pretransitional effects in DMPC-vesicle membranes: optical
  dynamometry study.
\newblock \emph{Biophys. J.} 79:340--356.

\bibitem[Kakorin et~al.(2003)Kakorin, Liese, and Neumann]{Kakorin:2003}
Kakorin, S., T.~Liese, and E.~Neumann, 2003.
\newblock Membrane curvature and high-field electroporation of lipid bilayer
  vesicles.
\newblock \emph{J. Phys. Chem.B} 107:10243--10251.

\bibitem[Torza et~al.(1971)Torza, Cox, and Mason]{Torza}
Torza, S., R.~Cox, and S.~Mason, 1971.
\newblock Electrohydrodynamic deformation and burst of liquid drops.
\newblock \emph{Phil. Trans. Royal Soc. A} 269:295--319.

\bibitem[Vizika and Saville(1992)]{Vizika:1992}
Vizika, O., and D.~A. Saville, 1992.
\newblock The electrohydrodynamic deformation of drops suspended in liquids in
  steady and oscillatory electric fields.
\newblock \emph{J. Fluid Mech.} 239:1--21.

\bibitem[Varshalovich et~al.(1988)Varshalovich, Moskalev, and
  Kheronskii]{Varshalovich:1988}
Varshalovich, D.~A., A.~N. Moskalev, and V.~K. Kheronskii, 1988.
\newblock Quantum Theory of Angular Momentum.
\newblock World Scientfic, Singapore.

\bibitem[Ha and Yang(2000)]{Ha:2000c}
Ha, J.~W., and S.~M. Yang, 2000.
\newblock Electrohydrodynamic effects on the deformation and orientation of a
  liquid capsule in a linear flow.
\newblock \emph{Phys. Fluids} 12:1671--1684.

\bibitem[Misbah(2006)]{Misbah:2006}
Misbah, C., 2006.
\newblock Vacillating breathing and tumbling of vesicles under shear flow.
\newblock \emph{Phys. Rev. Lett.} 96:028104.

\bibitem[Lebedev et~al.(2007)Lebedev, Turitsyn, and Vergeles]{Lebedev:2008}
Lebedev, V.~V., K.~S. Turitsyn, and S.~S. Vergeles, 2007.
\newblock Dynamics of nearly spherical vesicles in an external flow.
\newblock \emph{Phys. Rev. Lett.} 99:218101.

\end{thebibliography}
\clearpage

\centerline{\bf{FIGURE LEGENDS}}
\vspace{1cm}

\begin{figure}
\centerline{\includegraphics[width=4.5in]{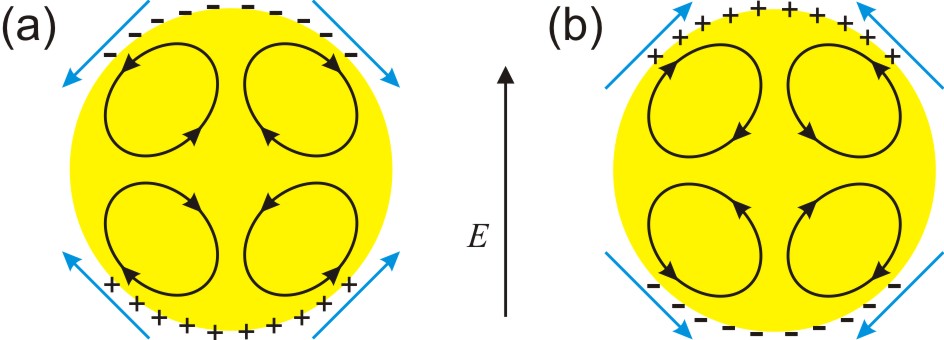}}
\caption{}
\label{fig: toroid}
\end{figure}

{\bf Figure \ref{fig: toroid}}: An illustration of the streamlines of the electrohydrodynamic flow inside a drop, surface charge distribution. The corresponding direction of the tangential electric traction is denoted by arrows  \cite{Taylor:1966, Torza}. (a) interior fluid less conducting than the exterior one,  $\Lambda/S<1$; (b) interior fluid more conducting than the exterior one, $\Lambda/S>1$.

\begin{figure}
\centerline{\includegraphics[width=4.5in]{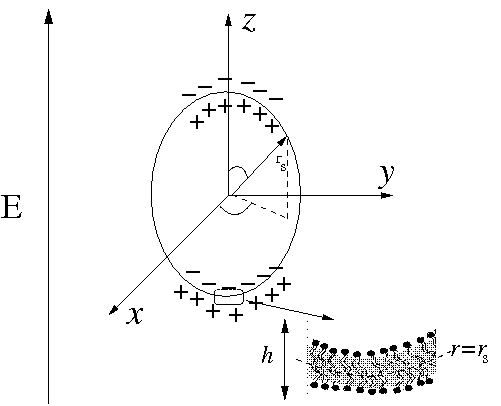}}
\begin{picture}(0,0)(0,0)
\put(-170,160){$\theta$}
\put(-180,110){$\phi$}
\end{picture}
\caption{ }
\label{sketch}
\end{figure}

{\bf {Figure \ref{sketch}}}: A sketch of a vesicle in a uniform electric field. The zoomed region of the interface illustrates the lipid bilayer  structure of the membrane.

\begin{figure}[h]
\centerline{\includegraphics[width=4.5in]{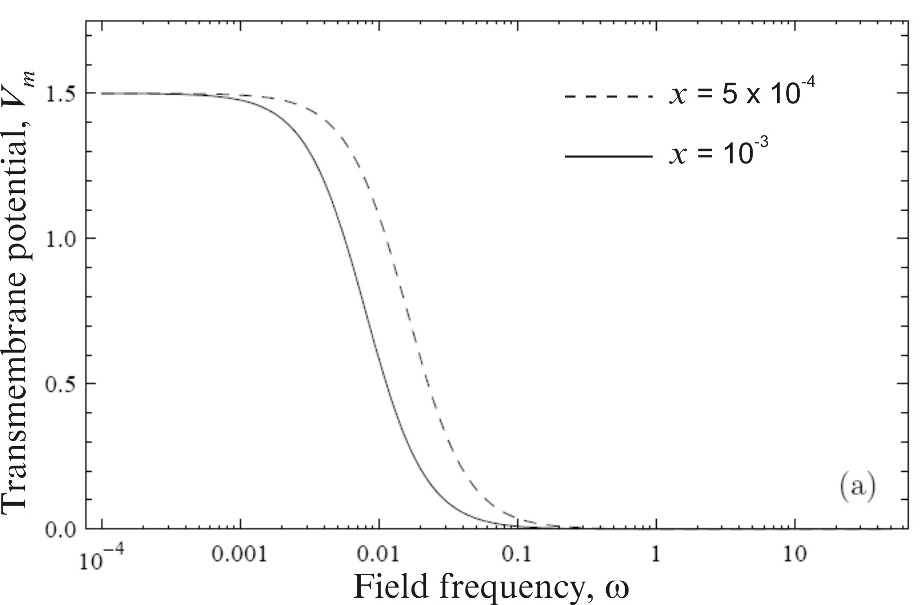}}
\centerline{\includegraphics[width=4.5in]{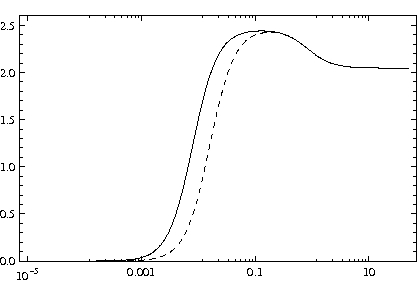}}
\begin{picture}(0,0)(0,0)
\put(50,30){\rotatebox{90}{inner electric field $Re[P^\ins]$}}
\put(230,0){field frequency $\omega$}
\put(360,50){\large{(b)}}
\end{picture}
\caption{ }
\label{fig:potential}
\end{figure}

{\bf Figure \ref{fig:potential}}:  (a) The transmembrane potential at the poles $\theta=0,\pi$ calculated from \refeq{potential3},  and
 (b) the interior electric field calculated from \refeq{el field2} for a spherical shell with membrane thickness  $x=5\times 10^{-4}$ (solid line) and $x=10^{-3}$ (dashed line) in a uniform AC electric field. $\Lambda=0.5,\, S=1.001, \,C_m=0.025/x,\, G_m=0$.

{\bf Figure \ref{fig:tractions}}: Electric tractions as a function of frequency for conductivity ratios  $\Lambda=0.5$ (solid line) and $\Lambda=1.5$ (dashed line). The other parameters are  $S=1.001, C_m=0.025/x, G_m=0$ and membrane  thickness $x=5\times 10^{-4}$. Dotted lines represent the electric tractions in the case of a droplet (the zero-membrane-thickness  limit).\\
(a) Electric pressure calculated from \refeq{tn}. \\
(b) Tangential electric force calculated from \refeq{tt}.

{\bf Figure \ref{fig:charge}}: The absolute value of the amplitude of the interfacial charge density at the poles $\theta=0, \pi$. Parameters are as in Fig.~\ref{fig:tractions}; only $\Lambda=0.5$ is plotted. The dashed line represents the induced charge in the case of a droplet (the zero-membrane-thickness  limit).

{\bf Figure \ref{shapef20}}: Evolution of the ellipsoidal deformation $f_{20}$, calculated from \refeq{f20AC-1},
of  a quasi-spherical vesicle upon application of a uniform AC electric field. Parameter values are $\Delta=0.2$, $\Lambda=1.5$,   $x=5\times 10^{-4}$, $G_m=0$, and $C_m=50$. $\Ca=684$, which corresponds to $E_0=10^4V/m$, $\eta_\out=10^{-3} Pa.s$,  $\kappa=25 k_BT$ and $a=10\mu m$.\\
(a) for frequencies $\omega=0.01,\, 0.1,\, 1$, denoted by solid, dashed and long-dashed lines,  viscosity ratio $\visrat=1$ and $\visrat_s=0$. \\
(b) for viscosity ratios $\visrat=0,\,1,\,5$, denoted by solid, dashed and long-dashed lines, at AC field frequency $\omega= 0.01$ and  $\visrat_s=0$.\\
(c) for membrane viscosity parameters $\visrat_s=0,\, 10,\, 100$  denoted by solid, dashed, and long-dashed lines, at AC field frequency $\omega= 0.01$ and viscosity ratio $\visrat=1$.

{\bf Figure \ref{shapesAC}}: {Ellipsoidal deformation $f_{20}/f^{max}_{20}$
for vesicles in AC field as a function of frequency. Solid line is for conductivity ratio $\Lambda=0.5$ and  dashed line is for $\Lambda=1.5$. Other parameters are $\Delta=0.2$,  $\visrat=1$, $\visrat_s=0$, $x=5\times 10^{-4}$, $G_m=0$ and $C_m=50$.}

{\bf Figure \ref{fig: trans freq}}: {Prolate-oblate transition frequency for different conductivity ratios $\Lambda$. The points are experimental data \cite{Aranda:2007, Dimova-Aranda:2007} for interior fluid conductivities of about $10^{-4} S/m$
averaged over number of vesicles with different size. The solid line is the theoretical prediction using $x=10^{-4}$ corresponding to vesicle radius $a=50\mu m$. The  bottom dashed line is calculated using $x=2\times 10^{-5}$ and the top dashed line is calculated using $x=5\times 10^{-4}$ .  The basic charging time is estimated to be  $t_c=10^{-7}s$.}

{\bf Figure \ref{dipole_comp}}: {The electric tractions according to our model (solid line), the drop model (short-dashed line), the effective dipole theory (long-dashed line) and only exterior field contribution in the effective dipole theory (dot-dashed line). (a) electric pressure (b) tangential tractions. Parameters are: $\Delta=0.5$, $\Lambda=1.5$,  $\visrat_s=0$, $x=5\times 10^{-4}$, $G_m=0$, $C_m=50$ and $\Ca=684$.}

\clearpage


\clearpage

\clearpage

\clearpage

\begin{figure}[h]
\centerline{\includegraphics[width=4.5in]{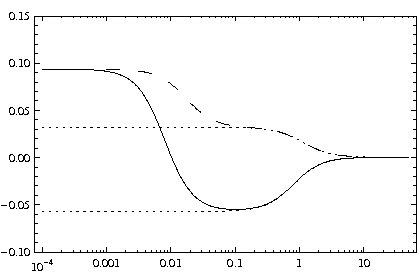}}
\begin{picture}(0,0)(0,0)
\put(50,30){\rotatebox{90}{electric pressure $\tau^{el}_r$}}
\put(230,0){field frequency $\omega$}
\put(360,50){{\large {(a)}}}
\put(300,200) {conductivity ratio}
\put(300,185){$\Lambda<1$ (solid line)}
\put(300,170) {$\Lambda>1$ (dashed line)}
\put(300,160) {$x=0$ (dotted line)}
\end{picture}
\end{figure}

\vspace{1cm}

\begin{figure}[h]
\centerline{\includegraphics[width=4.5in]{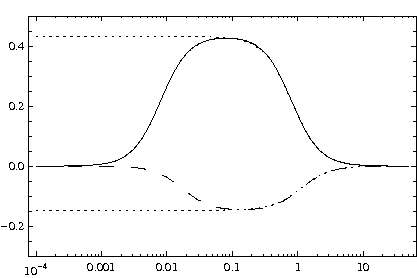}}
\begin{picture}(0,0)(0,0)
\put(50,30){\rotatebox{90}{tangential electric traction $\tau^{el}_\theta$}}
\put(230,0){field frequency $\omega$}
\put(360,50){{\large {(b)}}}
\end{picture}
\caption{ }
\label{fig:tractions}
\end{figure}
\clearpage

\begin{figure}
\centerline{\includegraphics[width=4.5in]{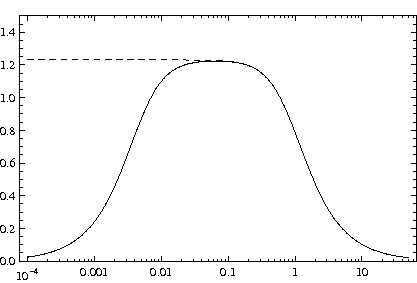}}
\begin{picture}(0,0)(0,0)
\put(50,30){\rotatebox{90}{surface electric charge $\bar Q$}}
\put(230,0){field frequency $\omega$}
\put(300,200) {membrane thickness}
\put(300,185){$x>0$ (solid line)}
\put(300,170) {$x=0$ (dashed line)}
\end{picture}
\caption{ }
\label{fig:charge}
\end{figure}
\clearpage

\begin{figure}[h]
\centerline{\includegraphics[width=4.5in]{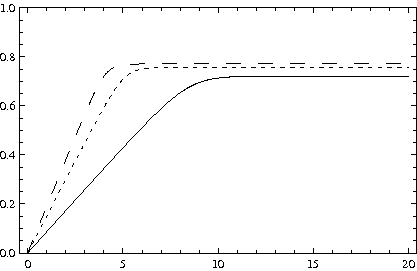}}
\begin{picture}(0,0)(0,0)
\put(50,30){\rotatebox{90}{shape elongation $f_{20}/f^{max}_{20}$}}
\put(230,0){dimensionless time $t/t_\el$}
\put(360,50){{\large {(a)}}}
\put(270,120) {field frequency:}
\put(270,105){$\omega=0.01$ (solid line)}
\put(270,90) {$\omega=0.1$ (short-dashed line)}
\put(270,75) {$\omega=1$ (long-dashed line)}
\end{picture}
\end{figure}
\begin{figure}[h]
\centerline{\includegraphics[width=4.5in]{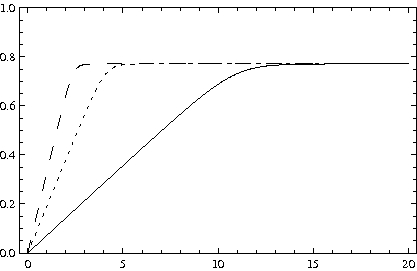}}
\begin{picture}(0,0)(0,0)
\put(50,30){\rotatebox{90}{shape elongation $f_{20}/f^{max}_{20}$}}
\put(230,0){dimensionless time $t/t_\el$}
\put(360,50){{\large {(b)}}}
\put(270,120) {viscosity ratio:}
\put(270,105){$\chi=0$ (solid line)}
\put(270,90) {$\chi=1$ (short-dashed line)}
\put(270,75) {$\chi=5$ (long-dashed line)}
\end{picture}
\end{figure}
\begin{figure}[h]
\centerline{\includegraphics[width=4.5in]{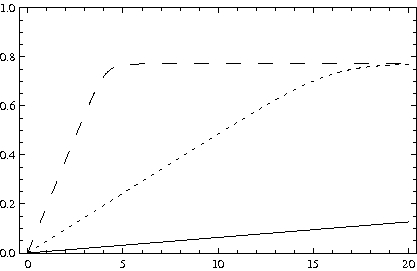}}
\begin{picture}(0,0)(0,0)
\put(50,30){\rotatebox{90}{shape elongation $f_{20}/f^{max}_{20}$}}
\put(230,0){dimensionless time $t/t_\el$}
\put(360,50){{\large {(c)}}}
\put(270,120) {membrane viscosity:}
\put(270,105){$\chi_s=0$ (solid line)}
\put(270,90) {$\chi_s=10$ (short-dashed line)}
\put(270,75) {$\chi_s=100$ (long-dashed line)}
\end{picture}
\caption{ }
\label{shapef20}
\end{figure}

\clearpage


\begin{figure}
\centerline{\includegraphics[width=4.5in]{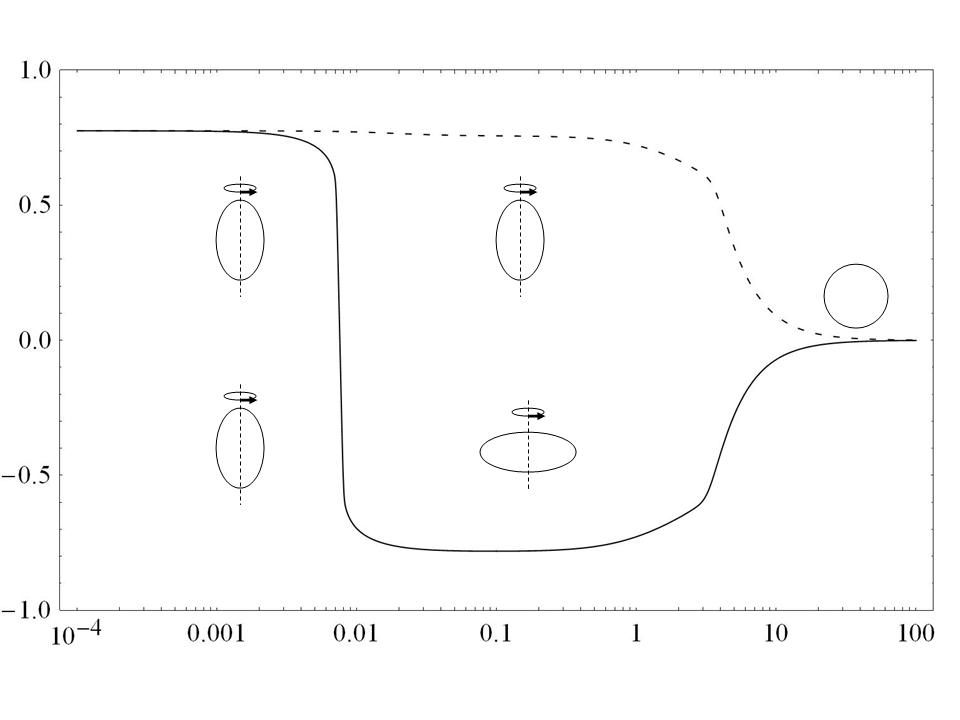}}
\begin{picture}(0,0)(0,0)
\put(50,80){\rotatebox{90}{shape elongation $f_{20}/f^{max}_{20}$}}
\put(200,10){field frequency $\omega$}
\end{picture}
\caption{ }
\label{shapesAC}
\end{figure}

\clearpage

\begin{figure}
\centerline{\includegraphics[width=4.5in]{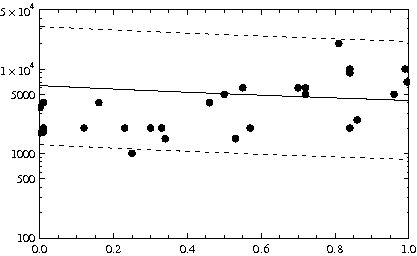}}
\begin{picture}(0,0)(0,0)
\put(50,30){\rotatebox{90}{conductivity ratio $\Lambda$}}
\put(230,0){field frequency $\omega/2\pi\, (Hz)$}
\end{picture}
\caption{ }
\label{fig: trans freq}
\end{figure}

\clearpage

\begin{figure}[h]
\centerline{\includegraphics[width=4.5in]{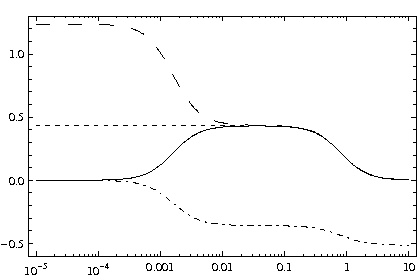}}
\begin{picture}(0,0)(0,0)
\put(50,30){\rotatebox{90}{electric pressure $\tau^{el}_r$}}
\put(230,0){field frequency $\omega$}
\put(360,50){{\large {(a)}}}
\end{picture}
\end{figure}
\begin{figure}[h]
\centerline{\includegraphics[width=4.5in]{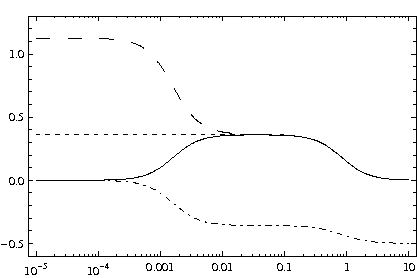}}
\begin{picture}(0,0)(0,0)
\put(50,30){\rotatebox{90}{tangential electric traction $\tau^{el}_\theta$}}
\put(230,0){field frequency $\omega$}
\put(360,50){{\large {(b)}}}
\end{picture}
\caption{ }
\label{dipole_comp}
\end{figure}

\clearpage

\end{document}